\documentclass[preprint]{emulateapj}

\newcommand{\myemail}{cmancone@astro.ufl.edu}

\newcommand{\sersic}{S\'{e}rsic}
\newcommand{\galfit}{GALFIT}
\newcommand{\pygfit}{PyGFit}
\newcommand{\pygsti}{PyGSTI}
\newcommand{\extractor}{Source Extractor}
\newcommand{\galapagos}{GALAPAGOS}
\newcommand{\tfit}{TFIT}
\newcommand{\convphot}{ConvPhot}
\newcommand{\spitzer}{{\it Spitzer}}
\newcommand{\hst}{{\it HST}}

\slugcomment{Accepted to PASP}

\shorttitle{PyGFit: The Python Galaxy Fitter}
\shortauthors{Mancone, Gonzalez, Moustakas, and Price}

\begin{document}


\title{PyGFit: A Tool For Extracting PSF Matched Photometry}

\author{Conor L. Mancone\altaffilmark{1},
Anthony H. Gonzalez\altaffilmark{1},
Leonidas A. Moustakas\altaffilmark{2},
Andrew Price\altaffilmark{2}
}
\altaffiltext{1}{Department of Astronomy, University of Florida, Gainesville, FL 32611}
\altaffiltext{2}{Jet Propulsion Laboratory, California Institute of Technology, 4800 Oak Grove Drive, Pasadena, CA 91109}

\email{\myemail}

\begin{abstract}
We present \pygfit{}, a program designed to measure PSF-matched photometry from images with disparate pixel scales and PSF sizes.  While \pygfit{} has a number of uses, its primary purpose is to extract robust spectral energy distributions (SEDs) from crowded images.  It does this by fitting blended sources in crowded, low resolution images with models generated from a higher resolution image.  This approach minimizes the impact of crowding and also yields consistently measured fluxes in different filters, minimizing systematic uncertainty in the final SEDs.  
We present an example of applying \pygfit{} to real data and perform simulations to test its fidelity.  
The uncertainty in the best-fit flux rises sharply as a function of nearest-neighbor distance for objects with a neighbor within $60\%$ of the PSF size.  Similarly, the uncertainty increases quickly for objects blended with a neighbor $\gtrsim4$ times brighter.  For all other objects the fidelity of \pygfit{}'s results depends only on flux, and the uncertainty is primarily limited by sky noise.
\end{abstract}

\keywords{Data Analysis and Techniques}

\section{Introduction}\label{sec:intro}

Astronomy has increasingly benefited from high-resolution imaging, exemplified by the Hubble Space Telescope, with a point-spread function (PSF) size of $<0.1^{\prime\prime}$ \citep{wfc3handbook}.  Obtaining ancillary multiwavelength data at comparable resolution is often impractical, and it is commonly necessary to work with mixed resolution data sets.  For instance, lower resolution ground-based data are often used in conjunction with high resolution space-based imaging, and at infrared wavelengths higher-resolution imaging is often not available or feasible.  In such cases, effective crowding can vary substantially as a function of wavelength, and the quality of the final data set is limited by the reliability of fluxes extracted from the most crowded images.  So long as sources remain unresolved, PSF fitting provides a viable method for extracting fluxes in crowded fields.  However, a different procedure is needed to measure magnitudes of resolved or marginally-resolved sources in crowded fields.  With mixed-resolution datasets, such a procedure must measure magnitudes in a consistent way despite differences in PSF, resolution, and crowding.

We present a new program, Python Galaxy Fitter (\pygfit{}){\footnote{\url{http://www.baryons.org/pygfit}}} aimed at solving these problems.  \pygfit{} is not the first program to address these issues (see for example \citealt{soto99}, \citealt{labbe05}, \citealt{tfit}, and \citealt{desantis07}).  Indeed, \pygfit{} and the well-known codes \tfit{} \citep{tfit} and \convphot{} \citep{desantis07} are conceptually similar.  \tfit{} and \convphot{} work by taking cutouts from a high-resolution image (HRI), convolving them with the low-resolution PSF, and fitting these models directly to the low-resolution image (LRI).  As a result, the HRI and LRI must be astrometrically aligned, and their pixels have to be properly matched.  In the case of \convphot{} the pixel scale must be the same in the low- and high-resolution images, and any offsets between the images must be integer pixel offsets and have to be fed into the program.  In the case of \tfit{}, the pixel scale of the LRI must be an integer factor of the pixel scale of the HRI, and the images must cover exactly the same area on the sky.  In both cases any sub-pixel offsets between the two images can introduce errors into the final results \citep{tfit}.

\pygfit{}, however, uses analytic source models.  It works on the basis of a high-resolution catalog (HRC) which gives the parameters of a model fit (i.e. a \sersic{} profile) for every object in the HRI.  \pygfit{} fits those models to the LRI, simultaneously fitting blended sources.  The use of models minimizes the impact of shot-noise in the HRI, especially for objects with low S/N ratio.  Also, this decouples the HRI and LRI, such that the LRI can have an arbitrary pixel scale and can cover larger or smaller areas on the sky.  Surveys with high resolution imaging routinely fit model profiles to all visible sources, which means that \pygfit{} can often build off of already existing catalogs.  Moreover, \pygfit{} is relatively fast, in many cases taking just a few minutes to fit the LRI for an area of the sky corresponding to a single {\itshape HST} pointing.  \pygfit{} performs an alignment step to account for any zeroth order offsets between the WCS of the HRI and LRI, and can also account for small sub-pixel shifts between the two images, which may arise due to either small imprefections in the WCS solutions or morphological k-corrections that subtly shift the object centroid.  \pygfit{} currently supports two models, a point source and \sersic{} model, and is extensible to include any analytic profile.  It also has a built-in capability to quantify uncertainties via simulations of artificial galaxies.

The intended purpose of \pygfit{} is to measure PSF-matched photometry from mixed-resolution datasets, especially for marginally-resolved sources in crowded fields.  This enables reliable measurements of galaxy SEDs and consequently, stellar mass fits.  However, \pygfit{} is not limited to this single application.  As a profile fitting routine, it has a number of other potential uses.  For instance, \pygfit{} can be used to subtract foreground sources from an image to search for faint background sources (such as gravitational arcs).  It can also subtract objects identified in one image from another image (presumably taken at a different wavelength), a feature that can be used, for instance, to identify high-z dropout candidates.

This paper is structured as follows.  Section \ref{sec:procedure} describes \pygfit{}'s fitting procedure.  Section \ref{sec:example} demonstrates \pygfit{}'s usage on real data and discusses some relevant limitations.  Section \ref{sec:sims} describes the simulations built into \pygfit{} and uses them to measure the fidelity of \pygfit{}.  Finally, Section \ref{sec:conclusions} gives our conclusions.  All magnitudes are on the Vega system, and we assume a WMAP 7 cosmology (\citealt{komatsu11}; $\Omega_m=0.272, \Omega_\Lambda=0.728, h=0.704$) throughout.

\section{Procedure}\label{sec:procedure}

\subsection{Overview}

The fundamental goal of \pygfit{} is to enable matched photometry in mixed-resolution data sets that is robust to the effects of crowding in the lower resolution images. In the limiting case where a source is effectively a point source in the lower resolution data set, this problem has long been solved as it is effectively a matrix inversion process (see e.g. the MOPEX software for MIPS photometry; \citealt{mopex}, or DAOPHOT; \citealt{stetson87}). Optimal deblending however becomes more challenging and computationally intensive when sources are even marginally-resolved, and the convolution of the PSF and underlying galaxy profile must be considered. 

With \pygfit{} we present an approach that is designed to be fast, flexible, and reliable. This code was originally designed to enable such robust photometry in the crowded cores of high-redshift cluster galaxies using the combination of \hst\ and \spitzer\ data, but is generally applicable to any situation in which one desires profile-matched photometry between mixed resolution data sets. As described below, \pygfit{} can successfully deblend the photometry of two sources as long as their intrinsic separation is more than approximately $60\%$ of the FWHM of the PSF in the low resolution data.  It is also important to note that \pygfit{} makes the implicit assumption that the shape of the underlying profile is the same at all wavelengths $-$ effectively an assumption that morphological $k-$corrections are small. In cases where the morphology changes strongly with wavelength, such as a starburst galaxy with an underlying old stellar population, the results from \pygfit{} should be treated with care.  Such cases may be flagged through the galaxies' colors and SEDs.

At its core, \pygfit{} uses position and shape information of objects in a high-resolution image (HRI) to determine how to divide the luminosity of overlapping objects in a low-resolution image (LRI) among the constituent components.  As such, the primary input into \pygfit{} is a high-resolution catalog (HRC) that gives positions and shapes of all objects in the HRI.  \pygfit{}'s procedure can be broadly separated into four steps: object detection and segmentation of the LRI, alignment of the HRC with the LRI, object fitting, and final catalog generation.  There are five primary inputs into \pygfit{} which must be provided: the HRC, the LRI, an RMS map for the LRI, the PSF image of the LRI, and a \extractor{} configuration file for the LRI.

The HRC should give \sersic{} model parameters for all objects in the HRI which are resolved in the LRI.  This requires fitting a \sersic{} profile to every object in the HRI, a task which is becoming common for surveys with {\itshape HST} imaging.  While any program can be used to fit models to the HRI, the modeling routines built into \pygfit{} use precisely the same equations as \galfit{} \citep{peng02,peng10}, enabling the output from \galfit{} to be fed directly into \pygfit{}.  Therefore, the simplest way to build the HRC is by using a program that can run \galfit{} and fit a \sersic{} profile to every object in the image (for example \galapagos{}, \citealt{galapagos}).

The first step \pygfit{} executes, object detection and segmentation, is performed by running \extractor{} \citep{bertin96} on the LRI.  The primary goal of this step is to generate a segmentation map of the LRI.  This provides a convenient method for determining which objects in the HRC are blended together and hence must be modeled together, and it also divides the process into manageable chunks.  \extractor{} also creates a low-resolution catalog (LRC) and a background map.  \pygfit{} stores the LRC and includes any desired information from it in the final output catalog.  The background map is used to estimate the sky for all objects, and is subtracted from the LRI before fitting.

This is followed by an alignment step between the HRC and the LRI which serves two purposes.  First, it accounts for any zeroth order offsets between the WCS of the HRC and the LRI.  Next, it accounts for any miscentering of the low-resolution PSF image.  \pygfit{} performs this global alignment by finding isolated objects and calculating the offset via least squares minimization.  \pygfit{} takes the median best-fit position offset and then adjusts the positions of objects in the HRC accordingly.

\pygfit{} then moves on to fitting all the objects.  It iterates through the segmentation regions of the LRI (i.e. the low-resolution sources) and finds all overlapping objects from the HRC.  \pygfit{} then generates and stores a model for all matching objects from the HRC, convolves each with the low-resolution PSF as needed, and performs a $\chi^2$ minimization using a Levenberg$-$Marquardt algorithm to fit the models to the low-resolution source.  During the $\chi^2$ minimization only the positions and fluxes of the objects are left as free parameters.  All other \sersic{} parameters (radius, \sersic{} index, aspect ratio, and position angle) are held fixed.  The positions are restricted to small shifts (typically less than a pixel) and the fluxes are constrained to be positive.

\begin{figure*}
\epsscale{0.8}
\plotone{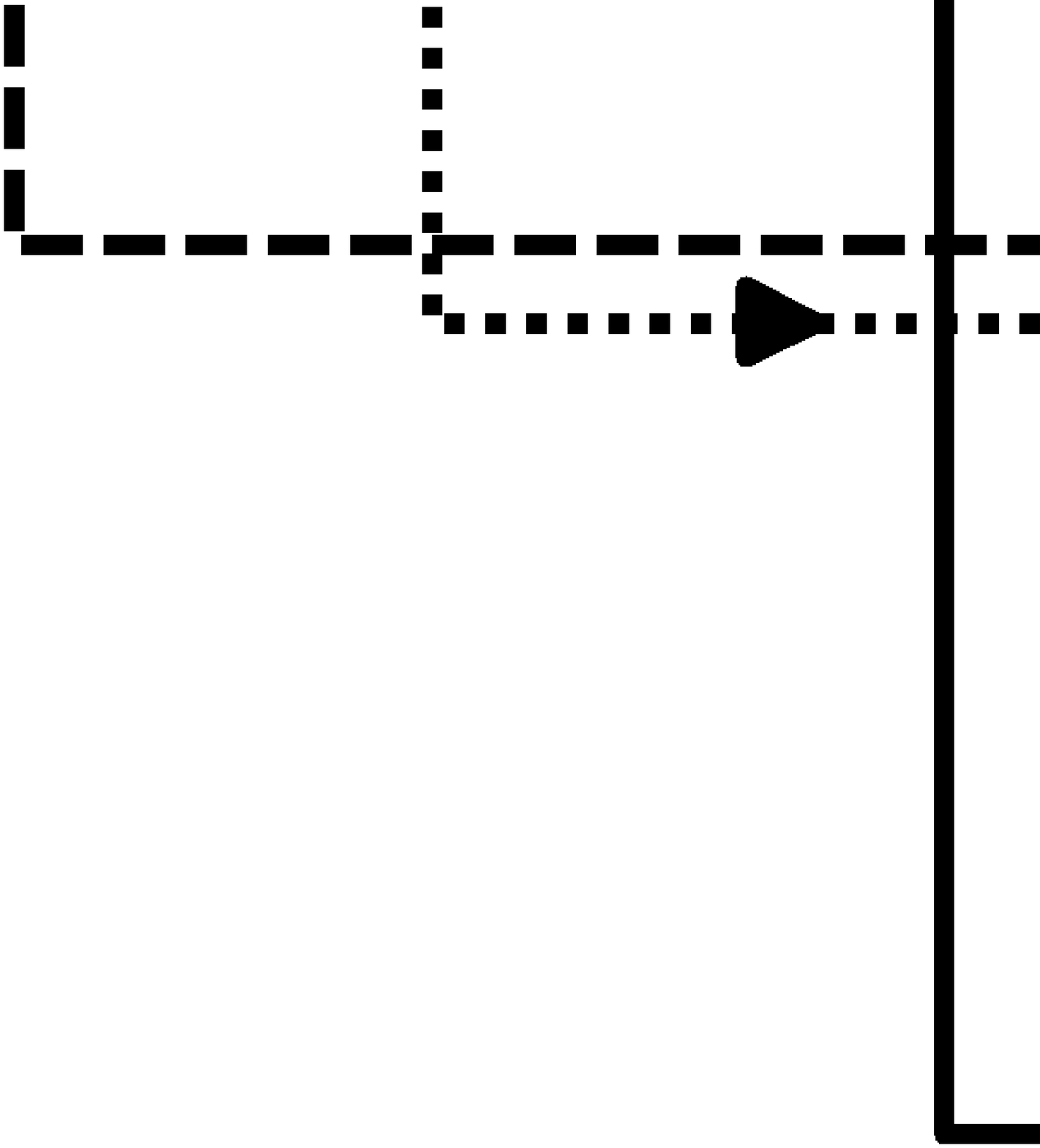}
\caption{A flow chart of \pygfit{}'s procedure.  Rectangles denote computational processes executed by \pygfit{} while the page symbols denote data products created by or used by \pygfit{}.  The five data products along the top are inputs which must be provided to \pygfit{}.}\label{fig:flow_chart}
\end{figure*}

Finally, the output catalog is generated.  This consists of the final fluxes measured by \pygfit{} for the objects in the HRC, any additional information requested from the LRC, and various diagnostics of each object.  At this stage \pygfit{} also generates a residual image for quick quality control and assessment.  Figure \ref{fig:flow_chart} gives a high level overview of \pygfit{}'s procedure, showing the primary inputs required by \pygfit{} on the top, the main steps it executes, and how the various inputs feed into each step.

\subsection{Object Detection and Segmentation}\label{sec:segmentation}

The first thing \pygfit{} does is to run \extractor{} on the low resolution.  It feeds the RMS map into \extractor{} and detection limits are set in a \extractor{} configuration file.  \pygfit{} uses three data products from this \extractor{} run: the segmentation map, the LRC, and the background map.  The most important output from \extractor{} is the segmentation map, which \pygfit{} uses to separate the process into distinct blocks .  We refer to each region of the segmentation map as a low resolution source.  A single low resolution source can have any number of objects from the HRC associated with it.  In practice, \pygfit{} ignores all low resolution sources which do not have any overlapping objects from the HRC.

One advantage of \extractor{} is that it has an easily configurable level of deblending.  It is preferable to minimize the amount of source deblending done by \extractor{} and rely on \pygfit{} to simultaneously fit the photometry for blended objects; 
however, 
a complete lack of deblending with \extractor{} can result in large segments of the LRI being assigned to one low resolution source.  These large segments in turn can cause unreasonably large execution times as \pygfit{} attempts to perform $\chi^2$ minimization for a problem with hundreds of free parameters.  In such cases allowing \extractor{} to perform a small amount of deblending can dramatically decrease execution with little to no loss of fidelity for the final results.

\extractor{} also generates a LRC which \pygfit{} stores.  \pygfit{} does not use any of the information in the LRC but simply passes it along to the final catalog.  By default, \pygfit{} extracts positions and auto-magnitudes from the LRC and copies them into the final catalog.  However, it can also pass along additional parameters from \extractor{} if desired.
Finally, \pygfit{} subtracts the background map from the LRI to remove its contribution from the low-resolution sources.

\subsection{Catalog Alignment}\label{sec:alignment}

Next \pygfit{} runs an alignment step.  The alignment step accounts for any zeroth-order misalignment of the HRC and LRI, as well as any miscentering of the low resolution PSF image.  \pygfit{} begins the alignment step by identifying isolated sources, i.e. low resolution sources which only have one associated object from the HRC.  \pygfit{} has configurable parameters to determine how many isolated sources should be used for the alignment step, as well as to limit them to a particular magnitude range.

After identifying isolated sources, \pygfit{} fits them using its normal routine (Section \ref{sec:fitting_procedure}) but with a larger allowed position shift than during normal fitting.  The precise size of the allowed position offset is configurable by the user, and should be large enough to account for any potential offset between the HRC and LRI.  Since there are only three free parameters being fit to the cutout from the LRI (x, y, and magnitude), there is no degeneracy and \pygfit{} easily recovers the position of the high resolution object in the LRI.  It is then straightforward to measure the median difference between the object positions in the HRC and the LRI and correct the HRC accordingly.  This step also accounts for any miscentering of the PSF template, which may happen when PSFs are determined empirically. If the PSF is not properly centered then the galaxy models will also be miscentered after PSF convolution.  The fitting process will naturally account for this offset, such that the final object positions will be shifted by the PSF offset.  Therefore when \pygfit{} performs the alignment step, it automatically corrects the HRC in such a way that the PSF-convolved galaxy models will be properly aligned with the LRI.

\subsection{Fitting}\label{sec:fitting_procedure}

\subsubsection{Cutouts}

The first step in the fitting process is to identify all low resolution sources which have matching objects from the aligned HRC.  Objects from the HRC are matched with a low resolution source if the object falls on one of the pixels identified by \extractor{} as belonging to the segmentation region for an object in the LRI.  Low resolution sources without any overlapping objects are ignored. Any desired further analysis of these sources can be performed separately using the residual image.  Fitting is done with the background-subtracted LRI, and fitting proceeds from the brightest low resolution sources to the faintest.  After each source is fit, the best-fit  model is subtracted from the LRI to remove its contribution to any nearby sources.

\pygfit{} generates a cutout of the blend from the LRI and extracts a matching cutout from the RMS map.  The extracted cutout is square and is large enough to enclose the full segmentation region of the low resolution source.  The cutout is further extended in every direction by the size of the allowed position shift during the fitting process, and an extra two pixels of buffer are added on each side.  If the resultant cutout image extends off of the LRI then \pygfit{} shifts the cutout to abut the edge of the image.

\begin{figure*}
\epsscale{1.0}
\plotone{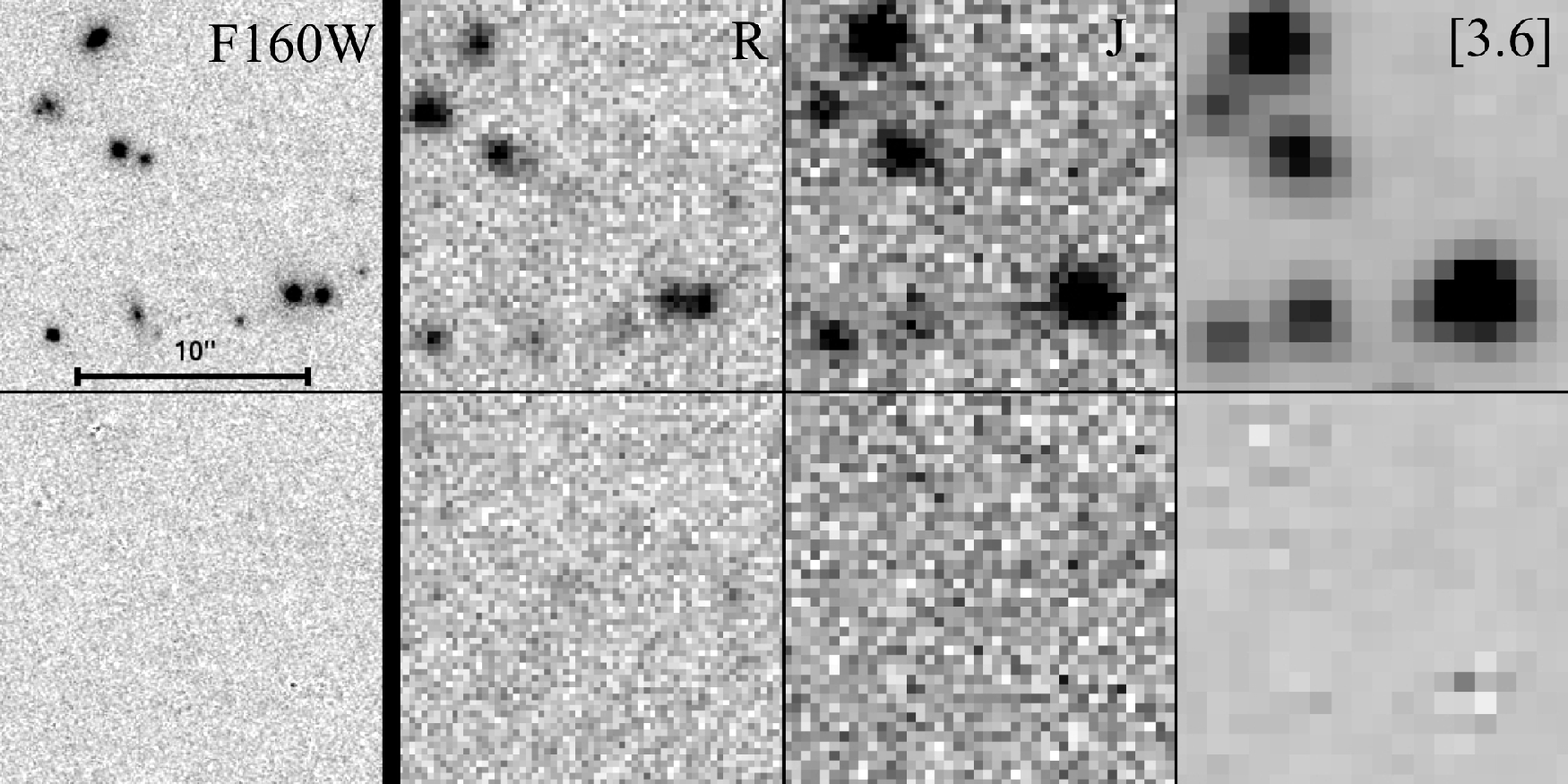}
\caption{Original images (top row) and residuals (bottom row) from our \galfit{} and \pygfit{} runs in the core of a high redshift ($z=1.243$) galaxy cluster.  From left to right the images correspond to WFC3/F160W, $R$, $H$, and $4.5\mu$m.  The WFC3/F160W image was fit with \galfit{}, while all other bands were fit with \pygfit{}.  All panels show the same field of view, and the scale in the top left panel is 10$^{\prime\prime}$ long. The ground-based images have seeing of 1$^{\prime\prime}$ ($R$) and 1.3$^{\prime\prime}$ ($H$), respectively.}\label{fig:residuals}
\end{figure*}

\subsubsection{Model Generation}\label{sec:modeling}

\pygfit{} then generates a model image for every matching object from the HRC. \pygfit{} currently supports two models, a point source and \sersic{} model, and is extensible to include any analytic profile. 
Model generation is very straightforward for point sources, which are simply a copy of the PSF image shifted to match the HRC position and scaled to match the total flux of the first guess used during the fit (Section \ref{sec:fitting}).  Shifting is accomplished with third-order spline interpolation.

\sersic{} model generation begins by calculating the average surface brightness ($\Sigma$) of the \sersic{} model in each pixel of the cutout.  The \sersic{} profile depends upon the effective radius ($r_e$), \sersic{} index ($n$), axis ratio ($B/A$), position angle (PA), total flux ($F_{tot}$), a boxiness parameter ($c$), and the profile center ($x_{cent}$, $y_{cent}$).  From these eight parameters \pygfit{} derives two more parameters: the surface brightness at the effective radius ($\Sigma_e$) and a coupling factor ($\kappa$) that ensures that the effective radius is also the half-light radius \citep[see for example][]{peng02}.  The surface brightness as a function of radius is then given by:

\begin{equation}\Sigma(r) = \Sigma_{e}e^{-\kappa[(r/r_e)^{1/n}-1]},\end{equation}

where

\begin{equation}\Sigma_e = \textrm{flux}*\textrm{R}(c) / [ 2{\pi}qr_e^2e^{\kappa}\kappa^{-2n}\Gamma(2n)n ]\end{equation}.

During model generation, \pygfit{} sets the flux according to its first guess for $\chi^2$ minimization (Section \ref{sec:fitting}).  $\Gamma(2n)$ is the gamma function and $\textrm{R}(c)$ is given by:

\begin{equation}\textrm{R}(c) = \frac{{\pi}c}{4\beta(1/c, 1+1/c)}\end{equation}

{\noindent}In this equation $\beta$ is the beta function with two parameters.  All of these definitions precisely match those for \galfit{} \citep{peng02,peng10}, which is done intentionally for ease of use.  If \galfit{} is used to fit \sersic{} profiles to the HRI, then the output from \galfit{} can be passed directly into \pygfit{} without modification.

\pygfit{} must calculate the flux in each pixel of the model image.  The most straightforward way to do this is to integrate the \sersic{} function over each pixel.  However, the integration time of the \sersic{} function can be computationally prohibitive, and \pygfit{} would be dramatically slower if it attempted to integrate the \sersic{} function over every pixel.  Instead \pygfit{} performs a numerical integration by splitting each pixel into subpixels, evaluating the \sersic{} function at each subpixel, and averaging their values together.  The level of resampling is finer towards the center of the model, with different levels of resampling for $r>2r_e$, $r<2r_e$, and the central pixel.  For these regions \pygfit{} resamples the model image such that the size of each subpixel is at most $r_e/2$, $r_e/20$, and $r_e/200$, respectively.

Extensive testing has shown that this methodology provides a reasonable execution time without compromising the results.  The only exception is for galaxies with small radii and high \sersic{} indexes ($n \sim 8$), where we find that the only way to reliably calculate the flux at the center of the \sersic{} profile is by directly calculating its integral.  However, these cases are easy to detect and, if desired, \pygfit{} can automatically switch from its default treatment to a full integration to guarantee that all galaxies are properly modeled.  After generating the \sersic{} model \pygfit{} then convolves it with the low-resolution PSF.

At the end of the model generation process \pygfit{} has a model image for every high-resolution object associated with a given low resolution source.  The generated model image matches the cutout for the blend.  The total flux of the model has been normalized to match the first guess that goes into the $\chi^2$ minimization (Section \ref{sec:fitting}), and the model has been convolved with the low-resolution PSF.  Therefore, all necessary steps have been performed to prepare the model images for fitting to the cutout of the low resolution source.

\subsubsection{Fitting}\label{sec:fitting}

\pygfit{} uses a Levenberg$-$Marquardt algorithm to minimize $\chi^2$ and fit the models to the low-resolution cutouts.  The cutout from the RMS image gives the uncertainty of every pixel in the cutout.  Each model has only three free parameters: $x$, $y$, and flux.  For \sersic{} models all other parameters ($n$, $r_e$, $B/A$, and PA) are held fixed to the values found in the HRC.  Since only the position and flux of the objects are free parameters, \pygfit{} does not need to generate a new model image at every iteration of the $\chi^2$ minimization.  Instead, \pygfit{} takes the stored model image, shifts it to match the new position (using third-order spline interpolation), and rescales it to match the new flux.  This is done for point source models as well as for \sersic{} models.  For each iteration of the $\chi^2$ minimization \pygfit{} takes the adjusted models, adds them together to make a total model image, and then calculates $\chi^2$ in the standard way.

The total number of free parameters ($n_f$) for each blend is given by $n_f=3*n_{HRC}$ where $n_{HRC}$ is the number of objects from the HRC associated with the low resolution source, and the total degrees of freedom for each fit is the number of pixels in the cutout minus the number of free parameters.  As a first guess for model positions \pygfit{} uses the object positions from the HRC after the alignment step.  The first guess value for the flux of each model is the magnitude of the object from the HRC converted to a flux using the zeropoint of the LRI.

During the fit the positions are constrained to move within a fixed distance of the first guess.  The size of the allowed position shift is easily configurable.  Rather than placing a constraint directly on the $\chi^2$ minimization, \pygfit{} uses a mapping function to convert the position offset calculated by the $\chi^2$ minimization from an infinite range to a finite range.  This keeps the positions within the desired offset without any modifications to the $\chi^2$  fitting routine.  We also force the fits to have positive fluxes.

\subsection{Final Catalog Output}\label{sec:catalog}

After fitting has been completed for all low resolution sources, \pygfit{} generates a final catalog.  The final catalog combines data from a number of sources.  It includes the best fitting magnitudes and fluxes for all matching objects in the HRC. The catalog also includes the object number and auto-magnitude for the low resolution source from \extractor{}, plus any other selected \extractor{} parameters.  All the information from the HRC is copied to the final output catalog.  Finally, \pygfit{} computes a number of diagnostic measures which can be included in the output catalog.  These include values such as the total number of high-resolution objects associated with the low-resolution source, the distance and best-fit magnitude of the nearest object in the blend, the best-fit magnitude of the brightest object in the blend, the total best-fit flux and magnitude of all objects in the blend, and the fraction of the blend flux which is accounted for by each object.

\section{Applying PyGFit to Real Data}\label{sec:example}

Our initial test case for \pygfit{} involved measuring SEDs of galaxies in high redshift galaxy clusters.  The data and project are described in detail in \citet{mancone13}.  In summary, we use 13 galaxy clusters with $1 < z < 1.9$ observed with broadband photometry in eleven filters.  All the clusters were observed with ground-based optical imaging in the B$_w$, R, and I bands, ground-based NIR imaging in the J, H, and K$_s$ bands, space-based NIR imaging in all four IRAC bands, and finally {\itshape HST} WFC3/F160W imaging.  We use \galapagos{} \citep{galapagos} to run \galfit{} and fit a single \sersic{} profile to every galaxy in our F160W images.  We then run \pygfit{} on each of the bands using the \galfit{} catalog as the HRC.

Figure \ref{fig:residuals} illustrates typical results from \pygfit{}.  It shows the original images in four different bands and their residuals after fitting in the center of ISCS J1434.5+3427, a galaxy cluster at $z=1.243$.  \pygfit{} cleanly subtracts all the visible objects.  This is very typical for our ground-based imaging, especially in the NIR where the residuals are indistinguishable from sky noise in virtually all cases.

\begin{figure}
\epsscale{1.0}
\plotone{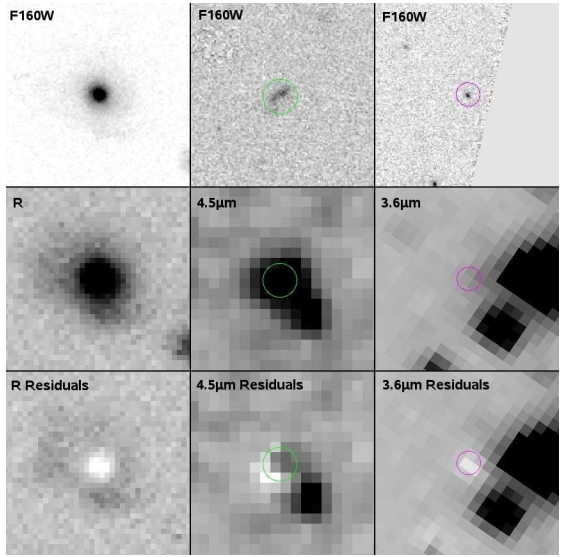}
\caption{Three examples of cases where \pygfit{} can fail.  The top row of panels shows the F160W images used to create the HRC.  The center row of panels shows the LRI with the same field of view, while the bottom row of panels shows the residuals of the LRI after fitting.  The LRI for the left uses ground-based R band imaging, while the example in the center is taken from the $4.5\mu$m imaging and the example on the right is taken from the IRAC $3.6\mu$m imaging.  The left column shows a galaxy with extended features which cannot be described by a \sersic{} profile.  The center column shows a galaxy which is isolated in F160W but which is blended with another source in $4.5\mu$m.  The right column shows a galaxy near the edge of the F160W image which is blended with a bright source which is outside of the F160W image.  Further details are in the text.}\label{fig:problems}
\end{figure}

At $4.5\mu$m (far right of Figure \ref{fig:residuals}) there are small residuals visible in the very centers of many objects.  These residuals are common to both our $3.6\mu$m and $4.5\mu$m filters but are not visible in the $5.8$ and $8.0\mu$m filters.  The primary difference between the IRAC images is the size of the PSF, which varies from $1.66^{\prime\prime}$ to $1.98^{\prime\prime}$ \citep{irac}. Our IRAC images have been dithered and resampled to have a pixel scale of $0.865^{\prime\prime}/$pixel. For this pixel scale the 3.6$\mu$m and 4.5$\mu$m PSFs are slightly undersampled, while the longer wavelengths are Nyquist sampled.

Without a fully resolved PSF, interpolation (which \pygfit{} performs during model generation and fitting) can introduce artifacts, and this is likely the source of the small residuals observed in our blue IRAC bands.  However, our simulations (Section \ref{sec:sims}) conclusively demonstrate that \pygfit{} can reliably extract magnitudes and fluxes from the observations, and that the primary source of uncertainty is simply sky noise.

\begin{figure}
\epsscale{1.2}
\plotone{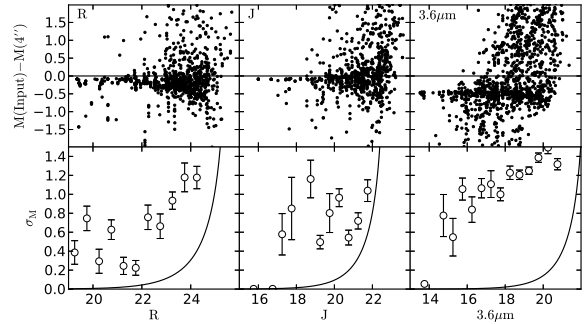}
\caption{(Top) The magnitude differences between measured 4$^{\prime\prime}$ aperture magnitudes and the input magitudes as measured for simulated galaxies in three filters: R (left), J (center), and $3.6\mu$m (right). No aperture corrections have been applied; the mean offsets from zero correspond to the necessary aperture corrections. (Bottom) Magnitude error as a function of magnitude.  The open circles show the standard deviation as a function of aperture magnitude, which is significantly affected by outliers. The solid line corresponds to the limit of Poisson sky noise in the absence of crowding for a 4$^{\prime\prime}$ aperture.}\label{fig:mag_sims_aper}
\end{figure}
An examination of our residuals images reveals a few classes of problems which can result in \pygfit{} failures.  We show a few examples of these cases in Figure \ref{fig:problems}.  One source of difficulty arises when a galaxy is not well represented by a \sersic{} function.  In the example in Figure \ref{fig:problems} (far left) a galaxy has extended features which cannot be modeled by a single \sersic{} profile.  As a result, the central region of the galaxy is over-subtracted, while the outer region is under-subtracted.  As long as the galaxy does not have a substantial amount of flux outside of the model radius, \pygfit{} can still return an approximately correct total flux.  If the galaxy does have substantial flux outside of the model radius, \pygfit{} will underestimate the total flux of the galaxy.  However, any error introduced by a mismatched model will be the same for all filters.  Therefore, when using \pygfit{} to measure SEDs, this class of problem can lead to an underestimated SED normalization but will not introduce any additional filter-to-filter uncertainty in the SED.

\pygfit{} can fail catastrophically when objects in the LRI are missing from the HRC.  If, in the LRI, an object in the HRC is blended with another object which is not in the HRC, then \pygfit{} will assign flux from the second object to the first, overestimating its flux.  This can happen in a number of ways, two of which are illustrated in Figure \ref{fig:problems}.  The top central panel of Figure \ref{fig:problems} shows a faint galaxy.  The center panel shows that, in the $4.5\mu$m image, there appears to be a significant elongation towards the bottom right, which cannot be accounted for from the F160W image.  After subtraction (bottom center) there appears to be an object left over below and to the right of the F160W source.  The only way to explain this is with the presence of an object which is bright in $4.5\mu$m but nearly invisible in F160W, and which happens to be blended with the object visible in F160W.  As a result, the object from the HRC is overfit to account for the flux from the additional low-resolution object, and therefore its fit is unreliable.

The right set of panels in Figure \ref{fig:problems} show the same class of problem in another context.  This shows what can happen when the LRI extends past the HRC.  The top right panel shows an object which is near the edge of the F160W image.  In the $3.6\mu$m image (center right) a bright object happens to be nearby but is just outside of the F160W field of view, and is therefore missing from the HRC.  Although this second object is outside of the F160W field of view, it is bright enough to contribute substantially to the flux near the object of interest.  As a result, \pygfit{} overestimates the $3.6\mu$m flux of the object which is in the HRC.  While this particular problem can likely occur for any image, we see it most commonly in our IRAC images.  This is because our IRAC images have the highest source densities and the largest PSF of all of our images, and this combination increases the likelihood of having such a blend.

\begin{figure}
\epsscale{1.2}
\plotone{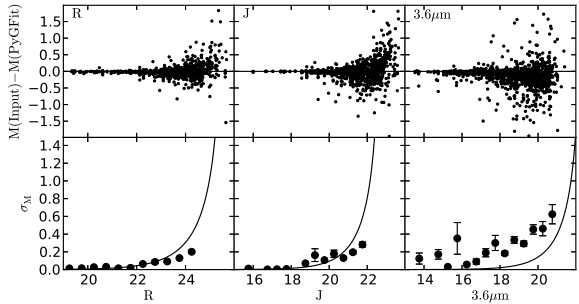}
\caption{(Top) The magnitude differences between \pygfit{} model magnitudes and the input magitudes as measured for simulated galaxies in three filters: R (left), J (center), and $3.6\mu$m (right).  (Bottom) Magnitude error as a function of magnitude.  The solid circles show the standard deviation as a function of aperture magnitude. For comparison, the solid line corresponds to the limit of Poisson sky noise in the absence of crowding for a 4$^{\prime\prime}$ aperture.}\label{fig:mag_sims_pygfit}
\end{figure}

Obviously, \pygfit{} cannot account for objects which are missing from the HRC.  This fact should be kept in mind when using \pygfit{} and care must be taken to include all sources which will be visible in the LRI, or to reject sources that are blended with objects missing from the HRC. The residual image generated by \pygfit{} can be used to identify drop-outs, and the $\chi^2$ statistics returned by the code can be used to identify objects that are poorly fit due to blends with objects that are missing in the HRC.

\section{Simulations}\label{sec:sims}

We use simulations to estimate the errors for the best-fit magnitudes and fluxes from \pygfit{}, as well as to evaluate its fidelity and limitations.  To aid in this process we developed a companion routine to \pygfit{} named \pygsti{} (Python Galaxy Simulation Tool for Images).  \pygsti{} uses the same model generation routines developed for \pygfit{}, generates simulated galaxies, and inserts them into images.  We have designed \pygfit{} to use \pygsti{} in a fully automated fashion.  We note that while \pygsti{} is packaged with \pygfit{}, it can also run as a stand-alone program and is convenient for generating simulated galaxies and stars for any number of applications.

\begin{figure}
\epsscale{1.2}
\plotone{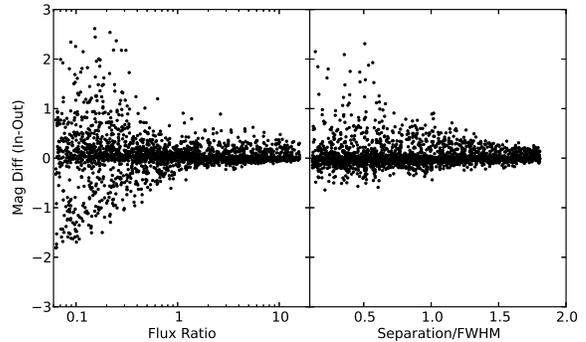}
\caption{Difference between input and output magnitude for simulated galaxies as a function of flux ratio (left) and separation relative to the size of the PSF (right).  To cleanly separate the two competing effects the left panel only includes galaxies separated by at least 1 PSF FWHM, and the right panel only includes galaxies which are the brightest galaxy in the pair.}\label{fig:close_sim}
\end{figure}

When running simulations, \pygfit{} randomly selects galaxies from the HRC, assigns them magnitudes from the magnitude distribution of the LRC, places them into random locations in the original image, runs \pygfit{} on the simulated frame, and repeats this process many times.  \pygfit{} limits the number of artificial galaxies placed into each simulated frame to prevent an excessive increase of crowding.  By default, the source density in \pygfit{}'s simulated frames is $2.5\%$ higher than in the original LRI, and \pygfit{} generates 100 simulated frames.  Both of these parameters are easily configurable.

Once \pygfit{} runs on all simulated frames, a final catalog is created with input and output magnitudes and fluxes for the simulated galaxies, along with all of the output parameters normally recorded by \pygfit{} (Section \ref{sec:catalog}).  This includes information on the number of objects the simulated galaxy was blended with, how close and bright the nearest neighbor is, and other environmental indicators.

\subsection{Simulation Results}

The simulations show that for a small percentage ($\sim2\%$) of galaxies, \pygfit{} dramatically underfits the model, effectively assigning them zero flux.  This occurs only for faint galaxies blended with brighter neighbors. In the simulations these galaxies are easily recognized as having a best-fit magnitude more than five magnitudes (100 times) fainter than the brightest galaxies in the blend. 
We find similar galaxies in our real data and note that they are easily detected by the same criteria.  Such galaxies should always be removed from any real sample as their magnitudes are completely unreliable.  Similarly, we remove them from our simulated galaxy sample and exclude them from further analysis.

We show the results of our simulations for three filters (R, J, and $3.6\mu$m) in Figures \ref{fig:mag_sims_aper} and \ref{fig:mag_sims_pygfit}, with the former showing results for aperture magnitudes and the latter results for \pygfit{}.  The top row of panels in each Figure shows the input and output magnitude of each simulated galaxy.  The bottom row of panels shows the corresponding error as a function of magnitude which is calculated by binning the simulated galaxies in magnitude space and measuring the standard deviation in each bin.  Error bars are calculated with bootstrap resampling.  The solid curve in the bottom row of panels shows an estimate of the magnitude error introduced by sky noise for an aperture magnitude with a diameter of $4^{\prime\prime}$.  
While the magnitudes returned by \pygfit{} are model fits rather than aperture magnitudes, this aperture provides a useful reference point for comparison because it is a common choice for \spitzer\ IRAC surveys \citep{eisenhardt2004,ashby2013}.

\begin{figure}
\epsscale{1.2}
\plotone{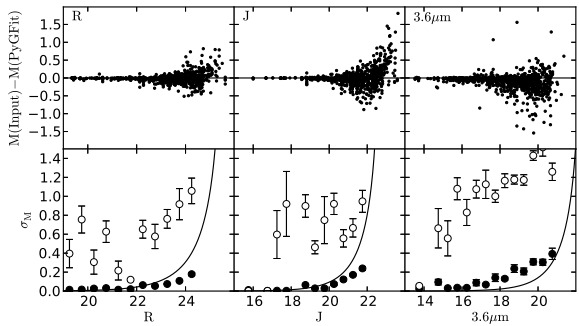}
\caption{Same as Figure \ref{fig:mag_sims_pygfit}, except simulated galaxies with separations $<60\%$ of the PSF radius or flux ratios $<0.25$ have been cut from the sample.}\label{fig:mag_sims_cut}
\end{figure}

For the $R$ and $J$ bands the magnitude errors that result from \pygfit{} are comparable to the expected sky noise for a $4^{\prime\prime}$ diameter aperture magnitude. These errors (solid circles in Figure \ref{fig:mag_sims_pygfit}) are substantially better than
for a simple $4^{\prime\prime}$ aperture (open circles in Figure \ref{fig:mag_sims_aper}), due to the impact of crowding on the aperture magnitudes. 
\pygfit{} does not however achieve performance comparable to the sky-noise limit for a $4^{\prime\prime}$ aperture magnitude in our $3.6\mu$m data.  A close examination of the top right panel of Figure \ref{fig:mag_sims_pygfit} shows that there are poorly fit galaxies ($|\textrm{M (Input)} - \textrm{M (PyGFit)}| > 0.75$) driving this scatter.  Our simulations reveal that \pygfit{} begins to break down when two galaxies are very close together or when a galaxy is blended with a much brighter one.  To quantify this, we perform another simulation where we insert pairs of galaxies into an image with the same noise properties, pixel scale, and PSF as the $3.6\mu$m image.  These simulated pairs have separations between $0.2^{\prime\prime}$ and $3^{\prime\prime}$, magnitude differences between 0 and 3 (i.e. flux ratios between 1 and $\sim15$), and the brighter galaxy in the pair has a magnitude between 15 and 17.  We drop these pairs into an otherwise blank image and measure \pygfit{}'s fidelity as a function of flux ratio and separation for close pairs.

Figure \ref{fig:close_sim} illustrates the result.  The left panel shows the magnitude error for simulated galaxies as a function of the flux ratio of a galaxy and its neighbors.  To isolate the influence of the flux ratio, this panel excludes galaxies separated by less than the FWHM of the PSF (1.66$^{\prime\prime}$ in $3.6\mu$m).  The right panel shows the magnitude error for simulated galaxies as a function of the separation between the pair relative to the FWHM of the PSF.  This panel only shows galaxies which are the brightest galaxy in the pair.  We find that our $3.6\mu$m \pygfit{} results become unreliable for galaxies with flux ratios $<0.25$ (i.e 1.5 magnitudes fainter than the blended object) or separations $\lesssim60\%$ ($\lesssim1^{\prime\prime}$) of the PSF radius.  Tests show that our other filters encounter a similar issue for such pairs.  However, source density is by far the highest in our IRAC images.  Because the crowding is less of an issue in our other bands, these limits have a smaller impact in our real and simulated data for our non-IRAC bands.  We therefore remove from our sample simulated galaxies within $1^{\prime\prime}$ separation of another object, or within 3$^{\prime\prime}$ of another object that is $>1.5$ mag brighter, and plot in Figure \ref{fig:mag_sims_cut} the fidelity of \pygfit{}'s results for the remaining galaxies.

\begin{figure}
\epsscale{1.20}
\plotone{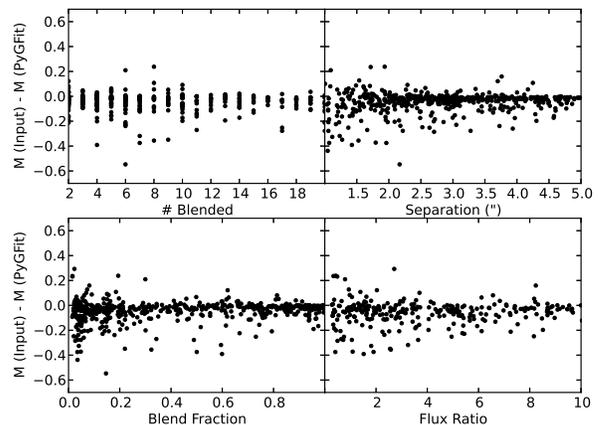}
\caption{\pygfit{} errors as measured with our simulations for our $3.6\mu$m galaxies with $[3.6] < 19.0$ versus the number of objects in the blend (top left), the distance to the nearest blended object (top right), the flux ratio between the object and its nearest neighbor (bottom right), and the fraction of the blend flux accounted for by the simulated object (bottom left).}\label{fig:ch1_error_correlations}
\end{figure}

Figure \ref{fig:mag_sims_cut} shows that after removing this problematic case of galaxies from our sample, the quality of the IRAC results is significantly improved. The standard deviation also decreases for the shorter wavelength bands, and is better than the $4^{\prime\prime}$ sky noise limit at faint magnitudes.  These simulations demonstrate that \pygfit{} provides reliable results fitting galaxies with separations as small as $\sim60$\% of the PSF FWHM, and in blends where the neighboring galaxy is less than 1.5 mag brighter. 

We examine how \pygfit{} performs as a function of environmental diagnostics.  In Figure \ref{fig:ch1_error_correlations} we show the fidelity of \pygfit{}'s results as a function of the number of objects blended together, the distance to the nearest object, the flux ratio between an object and its nearest neighbor, and the fraction of the blend accounted for by the simulated object.  We only plot simulated galaxies in this figure if they have $[3.6] < 19.0$ and pass the cuts discussed above (i.e. flux ratio $>0.25$ and separation $>1^{\prime\prime}$).  There are no strong correlations in Figure \ref{fig:ch1_error_correlations}, demonstrating that the quality of \pygfit{}'s results are independent of the degree of crowding or other environmental factors.  Similarly, we also find that there is no relationship between the uncertainty of \pygfit{}'s results and any of the \sersic{} parameters.  Other than magnitude, \pygfit{}'s fidelity is independent of $r_e$, $n$, $B/A$, and PA.

\section{Conclusions}\label{sec:conclusions}

We present \pygfit{}, a program which generates PSF-matched photometry from images with disparate pixel scales and PSF sizes.  
\pygfit{} takes model fits from high resolution images, fixes shape parameters, and fits the models to low resolution images allowing only the magnitudes to vary along with small position shifts.  
The code is publicly available at {\url http://www.baryons.org/pygfit}, along with additional documentation to facilitate use.
We apply \pygfit{} to real images and also perform simulations to measure \pygfit{}'s fidelity.  With the exception of some small residuals in the two bluest IRAC filters where the PSF is undersampled, \pygfit{} is able to cleanly subtract galaxies from the LRI.  Especially in the ground-based images, where the PSF is well resolved, only sky noise is visible in the residual images.
Simulations show that the uncertainty in \pygfit{}'s magnitudes are consistent with being limited by sky noise.

Our simulations identify a few classes of problems which can introduce errors into the \pygfit{} results.  Most important are catalog problems, i.e. incorrect or missing high resolution data.  Primary examples of catalog problems include fitting galaxy models in the HRI which are a poor fit to the galaxy, or the presence of objects in the LRI that are missing from the HRC.  The latter commonly happens because of differences in filter wavelength or because of the finite size of the HRI.
Our also simulations show that a small fraction ($\sim2\%$) of faint, blended galaxies are effectively ssigned zero flux.  While we find no obvious predictors for when this happens, such cases are rare and easy to detect/remove.  We further find that (as expected) \pygfit{} cannot deblend galaxies with arbitrarily close neighbors or arbitrarily bright companions.  This effect is important for our $3.6\mu$m data where crowding is the most prominent.  We find that \pygfit{}'s results are reliable down to separations as small as $\sim60\%$ of the PSF size.

\acknowledgements

The authors thank Chien Peng for helpful discussions at the outset of this project, and thank the referee for helpful suggestions ot improve the usability of the code.  The authors acknowledge support of this work from the National Science Foundation under grant AST-0070849.
The authors also acknowledge support through \hst programs 11597 and 11663, 
provided by NASA through a grant from the Space Telescope
Science Institute, which is operated by the Association of
Universities for Research in Astronomy, Inc., under NASA
contract NAS 5-26555.
The work of LAM was performed at the Jet Propulsion Laboratory, California Institute of Technology under a contract with NASA.

\bibliographystyle{apj}
\bibliography{pygfit}

\begin{thebibliography}{16}
\expandafter\ifx\csname natexlab\endcsname\relax\def\natexlab#1{#1}\fi

\bibitem[{{Ashby} {et~al.}(2013){Ashby}, {Stanford}, {Brodwin}, {Gonzalez},
  {Martinez}, {Bartlett}, {Benson}, {Bleem}, {Crawford}, {Dey}, {Dressler},
  {Eisenhardt}, {Galametz}, {Jannuzi}, {Marrone}, {Mei}, {Muzzin}, {Pacaud},
  {Pierre}, {Stern}, \& {Vieira}}]{ashby2013}
{Ashby}, M.~L.~N., {Stanford}, S.~A., {Brodwin}, M., {Gonzalez}, A.~H.,
  {Martinez}, J., {Bartlett}, J.~G., {Benson}, B.~A., {Bleem}, L.~E.,
  {Crawford}, T.~M., {Dey}, A., {Dressler}, A., {Eisenhardt}, P.~R.~M.,
  {Galametz}, A., {Jannuzi}, B.~T., {Marrone}, D.~P., {Mei}, S., {Muzzin}, A.,
  {Pacaud}, F., {Pierre}, M., {Stern}, D., \& {Vieira}, J.~D. 2013, ArXiv
  e-prints

\bibitem[{{Bertin} \& {Arnouts}(1996)}]{bertin96}
{Bertin}, E. \& {Arnouts}, S. 1996, \aaps, 117, 393

\bibitem[{{de Santis} {et~al.}(2007){de Santis}, {Grazian}, {Fontana}, \&
  {Santini}}]{desantis07}
{de Santis}, C., {Grazian}, A., {Fontana}, A., \& {Santini}, P. 2007, New
  Astronomy, 12, 271

\bibitem[{{Dressel}(2011)}]{wfc3handbook}
{Dressel}, L. 2011, {Wide Field Camera 3 Instrument Handbook, Version 4.0}
  (Baltimore, MD: STScI)

\bibitem[{{Eisenhardt} {et~al.}(2004){Eisenhardt}, {Stern}, {Brodwin}, {Fazio},
  {Rieke}, {Rieke}, {Werner}, {Wright}, {Allen}, {Arendt}, {Ashby}, {Barmby},
  {Forrest}, {Hora}, {Huang}, {Huchra}, {Pahre}, {Pipher}, {Reach}, {Smith},
  {Stauffer}, {Wang}, {Willner}, {Brown}, {Dey}, {Jannuzi}, \&
  {Tiede}}]{eisenhardt2004}
{Eisenhardt}, P.~R., {Stern}, D., {Brodwin}, M., {Fazio}, G.~G., {Rieke},
  G.~H., {Rieke}, M.~J., {Werner}, M.~W., {Wright}, E.~L., {Allen}, L.~E.,
  {Arendt}, R.~G., {Ashby}, M.~L.~N., {Barmby}, P., {Forrest}, W.~J., {Hora},
  J.~L., {Huang}, J.-S., {Huchra}, J., {Pahre}, M.~A., {Pipher}, J.~L.,
  {Reach}, W.~T., {Smith}, H.~A., {Stauffer}, J.~R., {Wang}, Z., {Willner},
  S.~P., {Brown}, M.~J.~I., {Dey}, A., {Jannuzi}, B.~T., \& {Tiede}, G.~P.
  2004, \apjs, 154, 48

\bibitem[{{Fazio} {et~al.}(2004){Fazio}, {Hora}, {Allen}, {Ashby}, {Barmby},
  {Deutsch}, {Huang}, {Kleiner}, {Marengo}, {Megeath}, {Melnick}, {Pahre},
  {Patten}, {Polizotti}, {Smith}, {Taylor}, {Wang}, {Willner}, {Hoffmann},
  {Pipher}, {Forrest}, {McMurty}, {McCreight}, {McKelvey}, {McMurray}, {Koch},
  {Moseley}, {Arendt}, {Mentzell}, {Marx}, {Losch}, {Mayman}, {Eichhorn},
  {Krebs}, {Jhabvala}, {Gezari}, {Fixsen}, {Flores}, {Shakoorzadeh}, {Jungo},
  {Hakun}, {Workman}, {Karpati}, {Kichak}, {Whitley}, {Mann}, {Tollestrup},
  {Eisenhardt}, {Stern}, {Gorjian}, {Bhattacharya}, {Carey}, {Nelson},
  {Glaccum}, {Lacy}, {Lowrance}, {Laine}, {Reach}, {Stauffer}, {Surace},
  {Wilson}, {Wright}, {Hoffman}, {Domingo}, \& {Cohen}}]{irac}
{Fazio}, G.~G., {Hora}, J.~L., {Allen}, L.~E., {Ashby}, M.~L.~N., {Barmby}, P.,
  {Deutsch}, L.~K., {Huang}, J.-S., {Kleiner}, S., {Marengo}, M., {Megeath},
  S.~T., {Melnick}, G.~J., {Pahre}, M.~A., {Patten}, B.~M., {Polizotti}, J.,
  {Smith}, H.~A., {Taylor}, R.~S., {Wang}, Z., {Willner}, S.~P., {Hoffmann},
  W.~F., {Pipher}, J.~L., {Forrest}, W.~J., {McMurty}, C.~W., {McCreight},
  C.~R., {McKelvey}, M.~E., {McMurray}, R.~E., {Koch}, D.~G., {Moseley}, S.~H.,
  {Arendt}, R.~G., {Mentzell}, J.~E., {Marx}, C.~T., {Losch}, P., {Mayman}, P.,
  {Eichhorn}, W., {Krebs}, D., {Jhabvala}, M., {Gezari}, D.~Y., {Fixsen},
  D.~J., {Flores}, J., {Shakoorzadeh}, K., {Jungo}, R., {Hakun}, C., {Workman},
  L., {Karpati}, G., {Kichak}, R., {Whitley}, R., {Mann}, S., {Tollestrup},
  E.~V., {Eisenhardt}, P., {Stern}, D., {Gorjian}, V., {Bhattacharya}, B.,
  {Carey}, S., {Nelson}, B.~O., {Glaccum}, W.~J., {Lacy}, M., {Lowrance},
  P.~J., {Laine}, S., {Reach}, W.~T., {Stauffer}, J.~A., {Surace}, J.~A.,
  {Wilson}, G., {Wright}, E.~L., {Hoffman}, A., {Domingo}, G., \& {Cohen}, M.
  2004, \apjs, 154, 10

\bibitem[{{Fern{\'a}ndez-Soto} {et~al.}(1999){Fern{\'a}ndez-Soto}, {Lanzetta},
  \& {Yahil}}]{soto99}
{Fern{\'a}ndez-Soto}, A., {Lanzetta}, K.~M., \& {Yahil}, A. 1999, \apj, 513, 34

\bibitem[{{H{\"a}u{\ss}ler} {et~al.}(2011){H{\"a}u{\ss}ler}, {Barden},
  {Bamford}, \& {Rojas}}]{galapagos}
{H{\"a}u{\ss}ler}, B., {Barden}, M., {Bamford}, S.~P., \& {Rojas}, A. 2011, in
  Astronomical Society of the Pacific Conference Series, Vol. 442, Astronomical
  Data Analysis Software and Systems XX, ed. I.~N. {Evans}, A.~{Accomazzi},
  D.~J. {Mink}, \& A.~H. {Rots}, 155

\bibitem[{{Komatsu} {et~al.}(2011){Komatsu}, {Smith}, {Dunkley}, {Bennett},
  {Gold}, {Hinshaw}, {Jarosik}, {Larson}, {Nolta}, {Page}, {Spergel},
  {Halpern}, {Hill}, {Kogut}, {Limon}, {Meyer}, {Odegard}, {Tucker}, {Weiland},
  {Wollack}, \& {Wright}}]{komatsu11}
{Komatsu}, E., {Smith}, K.~M., {Dunkley}, J., {Bennett}, C.~L., {Gold}, B.,
  {Hinshaw}, G., {Jarosik}, N., {Larson}, D., {Nolta}, M.~R., {Page}, L.,
  {Spergel}, D.~N., {Halpern}, M., {Hill}, R.~S., {Kogut}, A., {Limon}, M.,
  {Meyer}, S.~S., {Odegard}, N., {Tucker}, G.~S., {Weiland}, J.~L., {Wollack},
  E., \& {Wright}, E.~L. 2011, \apjs, 192, 18

\bibitem[{{Labb{\'e}} {et~al.}(2005){Labb{\'e}}, {Huang}, {Franx}, {Rudnick},
  {Barmby}, {Daddi}, {van Dokkum}, {Fazio}, {Schreiber}, {Moorwood}, {Rix},
  {R{\"o}ttgering}, {Trujillo}, \& {van der Werf}}]{labbe05}
{Labb{\'e}}, I., {Huang}, J., {Franx}, M., {Rudnick}, G., {Barmby}, P.,
  {Daddi}, E., {van Dokkum}, P.~G., {Fazio}, G.~G., {Schreiber}, N.~M.~F.,
  {Moorwood}, A.~F.~M., {Rix}, H.-W., {R{\"o}ttgering}, H., {Trujillo}, I., \&
  {van der Werf}, P. 2005, \apjl, 624, L81

\bibitem[{{Laidler} {et~al.}(2007){Laidler}, {Papovich}, {Grogin}, {Idzi},
  {Dickinson}, {Ferguson}, {Hilbert}, {Clubb}, \& {Ravindranath}}]{tfit}
{Laidler}, V.~G., {Papovich}, C., {Grogin}, N.~A., {Idzi}, R., {Dickinson}, M.,
  {Ferguson}, H.~C., {Hilbert}, B., {Clubb}, K., \& {Ravindranath}, S. 2007,
  \pasp, 119, 1325

\bibitem[{{Makovoz} \& {Marleau}(2005)}]{mopex}
{Makovoz}, D. \& {Marleau}, F.~R. 2005, \pasp, 117, 1113

\bibitem[{{Mancone et al.}(submitted)}]{mancone13}
{Mancone et al.}, C. submitted, \apj, 701, 428

\bibitem[{{Peng} {et~al.}(2002){Peng}, {Ho}, {Impey}, \& {Rix}}]{peng02}
{Peng}, C.~Y., {Ho}, L.~C., {Impey}, C.~D., \& {Rix}, H.-W. 2002, \aj, 124, 266

\bibitem[{{Peng} {et~al.}(2010){Peng}, {Ho}, {Impey}, \& {Rix}}]{peng10}
---. 2010, \aj, 139, 2097

\bibitem[{{Stetson}(1987)}]{stetson87}
{Stetson}, P.~B. 1987, \pasp, 99, 191

\end{thebibliography}

\end{document}